\documentclass[aps,prb,onecolumn,groupedaddress,floatfix,showpacs]{revtex4}
\usepackage{graphicx}
\begin{document}

\title{Embedded Boron Nitride Domains In Graphene Nanoribbons For Transport Gap Engineering.}

\author{Alejandro Lopez-Bezanilla$^{1,*}$}
\affiliation{$^1$ Oak Ridge National Laboratory, One Bethel Valley Road, Oak Ridge, Tennessee, 37831-6493, USA}
\author{Stephan Roche$^{2}$,$^{3}$}
\affiliation{
$^2$ CIN2 (ICN-CSIC) and Universitat Aut\'onoma de Barcelona, Catalan Institute of Nanotechnology, Campus UAB, 08193 Bellaterra (Barcelona), Spain\\
$^3$ ICREA, Instituci\'o Catalana de Recerca i Estudis Avancats, 08070 Barcelona, Spain
}
\footnote{*Author to whom correspondance should be addressed. alejandrolb@gmail.com}

\date{\today}

\begin{abstract}
We numerically investigate the impact of boron nitride (BN) domains on the transport properties of graphene
nanoribbons with lengths ranging from a few to several hundreds of nanometers and lateral size up to 4 nm. By
varying the size and morphology of the BN islands embedded in the graphene matrix, a wide transport tunability
is obtained from perfect insulating interfaces to asymmetric electron-hole transmission profiles, providing the
possibility to engineer mobility gaps to improve device performances. Even in the low-density limit of embedded
BN islands, transport properties are found to be highly dependent on both the BN-domain shape and the size with
a strong tendency toward an insulating regime when increasing the number of ionic bonds in the ribbon. This
versatility of conduction properties offers remarkable opportunities for transport gap engineering for the design
of complex device architectures based on a newly synthesized one-atom hybrid layered material.

\end{abstract}

% \pacs{72.80.Vp, 73.23.-b,71.15.Ap, 31.15.E-}


\maketitle

% \keywords{Keywords: Boron nitride, Graphene nanoribbon, Electronic transport, Heterostructures }
\section{\label{intro}INTRODUCTION}

Two-dimensional graphene has become a fantastic playground for exploring electronic and transport properties on the nanoscale.\cite{NovoselovRMP,PhysRevB.64.125428,PhysRevB.79.125421,springerlink:10.1007/s12274-008-8043-2} Among its many spectacular and unique features, charge transport has been found to be weakly sensitive to disorder, owing to quantum mechanisms and interference effects, such as the Klein tunneling and the weak antilocalization phenomenon \cite{ISI:000240766000023,PhysRevLett.97.146805,0295-5075-94-4-47006}, yielding anomalously large charge mobility up to room temperature \cite{ISI:000256641500001,ISI:000258325800014}. One limitation of the use of graphene in future nanoelectronics is the absence of an energy band gap, which prevents any efficient electronic current flow from switching off.\cite{ISI:000285574000004, PhysRevB.79.075407} The modification of graphene geometry and chemical structure with selective methods, such as the chemical attachment of functional groups with specific features, allows for mobility gap engineering. \cite{ISI:000297686500046,Li29022008,ISI:000271513600067} However, the covalent functionalization of graphene inevitably leads to the creation of $sp^3$ defects in the aromatic structure of graphene which eventually yields strong localization phenomena in graphene samples, reducing by orders of magnitude the otherwise large mobilities and associated current densities. 

In that perspective, a very appealing strategy to tune electronic features of graphene is to incorporate boron nitride (BN) domains inside the carbon matrix host to create a boron-nitride-carbon (BNC) heterostructure, taking advantage of the isomorphism between the graphene and the two-dimensional hexagonal BN layer, which exhibits a wide band gap on the order of 6 eV. \cite{PhysRevB.51.6868,PhysRevLett.96.026402,ISI:000221890700023,PhysRevLett.75.3918,ISI:000254669900075} The BN codoping of graphene materials introduces $sp^2$ bonds with a marked ionic character which breaks the $p_z$ network symmetry and disrupts the original graphene hyperconjugated $\pi$-network. Some impact on the electronic and transport properties is thus expected although to date not fully quantified. The successful experimental fabrication of atomic layers of hybridized patches of BN and graphene \cite{doi:10.1021/nl1022139},\cite{ISI:000276953500024} could pave the way for controlled tunability of electronic and transport properties. Indeed, a first transport experiment by Song {\it et al.} \cite{doi:10.1021/nl1022139} has revealed some anomalous temperature-dependent {\it insulating-to-metal transitions}, tentatively assigned to competing and parallel transport mechanisms. The reported mobilities of these hybrid monoatomic layers are limited to typically 10 ${\rm cm}^2{\rm V}^{-1}{\rm s}^{-1}$ with a resistance at the Dirac point of the order of $0.5{\rm M}\Omega$, suggesting that the random distribution of BN clusters and associated domain boundaries introduces a considerable source of scattering in the graphene matrix, despite the preservation of the hexagonal geometry. 

In this paper, we report a first-principles study of electronic quantum transport properties of disordered graphene nanoribbons (GNRs) modified with boron nitride domains (BNC-NRs). Using a combination of density-functional-theory-(DFT)-based calculations and simulations of mesoscopic electronic transport based on the Landauer-B\"uttiker (LB) approach, we unveil a broad spectrum of transport features in disordered graphene ribbons with embedded BN patches. We describe transmission profiles ranging from a total insulating behavior to asymmetric electron-hole conducting patterns and sequences of tunneling resonances in the presence of structural defects, such as vacancies inside the BN clusters. Various configurations of substitutional BN domains in graphene nanoribbons are presented, ranging from the co-doping of BN dimers at different sites across the nanoribbon width up to its complete coverage. This will allow us to explore different quantum transport mechanisms, including the quasiballistic, diffusive, or tunneling regimes. For hybrid BNC-NRs with initially large energy gaps, a superimposed BN-induced transport gap develops. The results are related to the BN-domain size and shape. We compare our findings with previous studies concerning the modification of the aromatic properties of GNRs upon chemical modification induced by functional groups. The richness of the obtained properties offers a wide electronic-transport tunability, which is suitable for engineering complex circuitry and device architecture based on such a two-dimensional platform.

\section{\label{methods}METHODS}
The self-consistent calculation of the electronic structure of a hybrid BNC system is performed with the localized orbital basis set of the SIESTA DFT-based code. \cite{PhysRevB.53.R10441,0953-8984-14-11-302} Thus a tight-binding-like Hamiltonian which presents the compactness required to study transport properties within the Green's function formalism is obtained. A double-$\zeta$ basis set with additional polarization orbitals is used to optimize the geometry of the hybrid ribbons and compute their electronic structures. The local density approximation (LDA) is adopted for the exchange correlation functional and the Troullier-Martins scheme \cite{PhysRevB.43.1993} is used to describe the interaction between ionic cores and localized pseudo-atomic orbitals. Hybrid BNC-NR are modeled within a supercell with an appropiate distance between neighboring cells to avoid spurious interactions. All atoms are relaxed with a force tolerance of 0.01 eV/\AA\  and the unit cell vectors are relaxed with the maximum stress component being smaller than 0.02 GPa. During geometry optimizations a 1x1x8 Monkhorst sampling is adopted in the Brillouin zone for 3.4 nm long ribbons. The numerical integrals are performed on a real space grid with an equivalent cutoff of 300 Ry. 

The electronic transport calculations are based on the LB formulation of conductance, which is particularly suitable to study the electron motion along a one-dimensional device channel in between two semi-infinite leads. We consider a phase coherent system formed by a scattering region where charge carriers can be backscattered during their propagation, and two semi-infinite leads in thermodynamical equilibrium with infinitely larger electron reservoirs. Within the LB approach, we calculate the transmission coefficients $T_n(E)$ for a given channel $n$ which gives the probability for an electron to be transmitted at a given energy $E$ when it quantum mechanically interferes with the BN domains. The conductance is expressed as $G(E)={G_0}\sum_{n}T_n(E)$, where $G_0$ is the quantum of conductance. The transmission coefficients are calculated
by evaluating retarded (advanced) Green's functions of the system, 

\begin{equation}
\mathcal{G}^{\pm}(\epsilon)=\{\epsilon I-H-\Sigma^{\pm}_{L}(\epsilon)-\Sigma^{\pm}_{R}(\epsilon)\}^{-1}
\end{equation}
where $\Sigma^{\pm}_{L(R)}(\epsilon)$ are the self energies describing the coupling of the channel to the left (right) lead. These quantities are related to the transmission factor by the Caroli's formula \cite{0022-3719-4-8-018}:
\begin{equation}
T(\epsilon)=tr\{\Gamma_{L}(\epsilon) \mathcal{G}^{+}(\epsilon) \Gamma_{R}(\epsilon) \mathcal{G}^{-}(\epsilon)\}
\end{equation}
with 
$\Gamma_{L(R)}(\epsilon)=i\{\Sigma^{+}_{L(R)}(\epsilon)-\Sigma^{-}_{L(R)}(\epsilon)\}$. $tr$ stands for the trace of the corresponding operator. 

To obtain the Green's function of long disordered ribbons,
we first calculate the Hamiltonian and overlap matrices of
short BNC-NR sections. These sections are then combined
in a random fashion with short pieces of pristine ribbon to
build up a system with disorder both along the ribbon length
and its lateral size. By means of real-space renormalization
techniques \cite{PhysRevLett.99.076803,ISI:000227408200009,doi:10.1021/nl802798q} we are able to evaluate the transport properties of BNC-NRs with lengths up to L=2$\mu$m. To avoid any artificial backscattering effects in the region where two ribbon segments match, the supercell in the {\it ab initio} calculation is chosen long enough to avoid any interference of the BN domains in the borders where periodic conditions are applied. 

\section{\label{Results}RESULTS}
In the following, we will refer to an armchair BNC-NR, composed of P -dimer lines, as P -aBNC-NR. Armchair GNRs are all semiconducting but with different energy band-gap widths. The pristine GNRs corresponding to the two BN-modified BNC-NRs studied here (14-aBNC-NR and 35-aBNC-NR) belong to the GNRs class with small energy band gaps.\cite{PhysRevLett.97.216803} Their widths are 1.6 nm and 4.1 nm respectively, which is within the reach of experiments. \cite{Li29022008,ISI:000271513600067} The 14-aBNC-NR supercells contain 256 atoms while the 35-aBNC-NR supercells contain up to 666 atoms. The dangling $sp^2$ $\sigma$-bonds of the edge atoms are passivated by H atoms. It is always observed that the BNC heterostructures preserve the flat geometry of the unmodified graphene independently of the relative $B_xN_yC_z$ composition although variations of the B-N, B-C and N-C bond lengths are observed for the various types of codopings. The insets in the upper panels of Figure \ref{FIG1} are a schematic representations of typical distances between neighbouring C, B, and N atoms, showing that, with respect to the usual 1.42 \AA\ of bulk graphene, N tends to shorten and B to enlarge the bond lenghts with neighbouring C atoms. This bond length variation is specially noticeable for atoms sitting on the edges. Experimental results in carbon nanotubes demonstrated that B and N are likely to codope into a carbon nanotube by substitution of a pair of carbon atoms, which results in BN existing as neighboring atomic pairs in the hexagonal structures.\cite{ADMA:ADMA200800830} According to the computed formation energy for the flat BN-codoped C structures tabulated in Table \ref{table1}, the configurations with B and N atoms in adjacent positions are all energetically more favourable than any of the considered geometries with B and N separated by one or two C atoms, as shown in intermediate panels of Figure \ref{FIG1}. Notice that the codoping with both or one of the BN atoms sitting on the edges is always energetically more favourable.

\begin{table}
  \begin{tabular}{llll}
    \hline
     & 	BN  & B-C-N & B-C-C-N  \\
    \hline
     i) & 0.71 & 2.00   &  2.08 \\
     ii) & 0.75 & 2.03   &  2.15 \\
     iii) & 0.72 & 1.85   &  1.93 \\
     iv) & 0    & 1.29   &  1.34 \\
  \end{tabular}

\caption{Relative formation energies for 14-aBNC-NRs with B and N atoms distributed across the ribbon width and at several relative distances, as shown in the ball-and-stick models in Figure \ref{FIG1}. BN denotes a boron nitride dimer, while B-C-N and B-C-C-N indicates that B and N atoms are sitting at second and third neighbours, respectively. Energies are compared to the most stable configuration and in units of eV.}
 \label{table1}
\end{table}

\begin{figure}[htp]
 \centering
 \includegraphics[width=0.45 \textwidth]{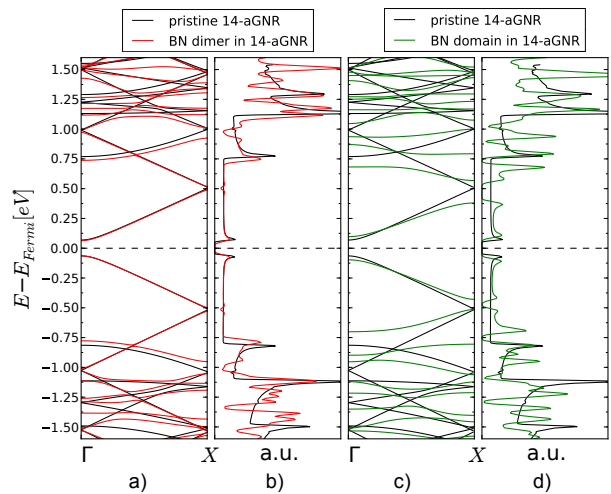}
 \caption{ Band stuctures (a and c) and total densities of states (b and d) of a pristine graphene ribbon (solid black lines in the four panels) and the 14-aBNC-NC with either a BN dimer (red solid lines, a and b), as shown in Figure \ref{FIG1}-a)-iii), or a BN domain (green solid lines, c and d), as shown in Figure \ref{FIG3}-a). Horizontal dashed lines indicate the Fermi energy level.}
\label{FIG0}
\end{figure}

For a better understanding of the effect of BN dimers and larger BN domains in the transport properties of GNR, it is worth comparing the electronic structures of the pristine GNR and the hybrid BNC ribbons. Figure \ref{FIG0} shows an energy band comparison of a pristine 14-aGNR with a 14-aBNC-NR which has a BN dimer located as shown in Figure \ref{FIG1}-a)-iii). The
presence of a BN pair does not introduce quasibound states,
and both the valence and the conduction bands are barely
affected by the presence of the impurity. On the contrary,
for a larger BN domain, as the one shown in the inset of
Figure \ref{FIG3}-a), the mixing of B and N electronic states with those
of the carbon matrix leads to symmetric small gap openings at
energies above $0.5$ eV and below $-0.5$ eV the Fermi level, and to a modification of the dispersiveness of the electronic
states. In the following, we study the consequence of these
electronic structure modifications for long GNRs, containing
a varied number of BN defects ranging from 5 up to 320.

\begin{figure*}[htp]
 \centering
 \includegraphics[width=0.95 \textwidth]{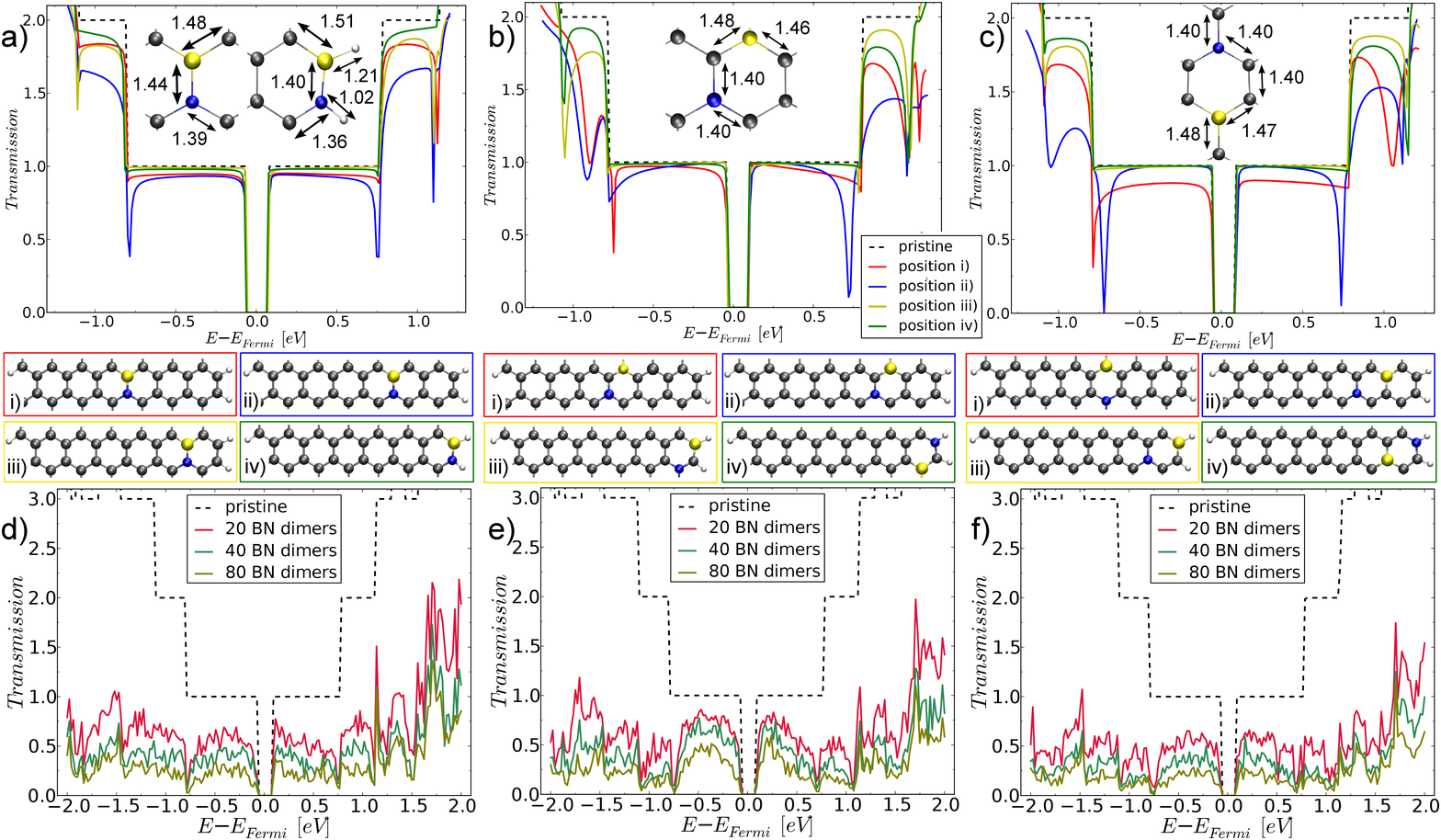}
 \caption{ Upper panels: Transmission coefficients for a 14-aBNC-NRs with a single BN pair located along the ribbon width and for varying distances between B and N atoms. a) gives T(E) for B and N first nearest neighbors as depicted in the inset and bottom ball and stick models. b) gives the same information for B and N second nearest neighbors. c) gives the same information for B and N third nearest neighbors. Dashed lines give the transmission profile for the pristine case. Bottom panels: Transmission profiles for $2\mu$m long disordered 14-aBNC-NRs, with a varying number of BN pairs, from 20 to 160.}
\label{FIG1}
\end{figure*}

Figure \ref{FIG1} (a,b,c) present a set of transmission profiles for the structures depicted by the corresponding ball-and-stick pictures [from (i) to (iv)] for four different BN pairs in substitutional positions across the ribbon width. The different cases
are defined by a varying distance between the doping atoms. The results are strongly position dependent, highlighting the
important interplay between co-doping site and edges. For the three types of co-doping, the configurations with one or
both B and N atoms sitting on the edge exhibit enhanced robustness against backscattering and, therefore, they are
likely to more efficiently preserve the resulting nanoribbon conductance in the presence of disorder. For two adjacent B
and N atoms [Figure \ref{FIG1}-a], the transmission coefficient drops are almost symmetrically positioned with respect to the Fermi
level as also found for BN-co-doped carbon nanotubes \cite{PhysRevB.81.193411} and are related to some Fano-resonance-effect coupling states
between different sub-bands (see Ref. \cite{ISI:000252018300001} for details). When B and N atoms are separated by one C atom [Figure \ref{FIG1}-b)] or sitting at opposite sites of a hexagon [Figure \ref{FIG1}-c)], an asymmetry of backscattering strength between hole and electron channels
is observed in the first plateau for some configurations and for all cases in the second plateau. Considering that edge
co-doping is energetically more favorable than bulk co-doping, in a controlled growth of BNC-NR, a prevalence of this type of
configurations is desirable to preserve charge-carrier mobility against disorder for long GNRs. Therefore, transmission
profiles of GNRs containing BN-pair co-doping exhibit a low drop and, in general, a symmetric shape at both sides of the
charge neutrality point.

Next, we study each case separately by considering systems
with many BN pairs. To simulate realistic systems, we consider
random distributions of BN dimers along the length and across
the width of 2-$\mu$m-long aBNC-NRs to analyze how the lack
of translational invariance will dictate the electronic-transport
regime. The mesoscopic transport calculations are performed
introducing 20, 40, and 80 BN dimers randomly distributed
along and across the nanoribbon, which represents about 99$\%$ in carbon over BN co-doping. Results are shown in panels
d)-f), respectively, of Figure \ref{FIG1}. It can be noticed that, for a
given number of BN pairs, the strength of backscattering is
significantly tuned by the separation between B and N dopants
from first to third neighboring positions. This is particularly
striking for the transmission profile with varying B and N
atoms lying at second-nearest neighbor positions [Figure \ref{FIG1}-f)] with respect to the others. In this situation, the electronic
transport in the first sub-bands clearly appears more stable
against BN co-doping. For all cases, the disorder has a larger
impact on high-energy sub-bands when compared with the
transmission profiles on the first sub-bands, which, even in the
low-defect density limit, always remain poorly conducting.
The observed behaviors for these sub-bands suggest that the
corresponding states are quickly driven to strong localization.
For the three types of co-doping and, in particular, for the
energetically most favorable cases with adjacent B's and N's,
an asymmetry between electron and hole conduction develops
at high energies. It is worth noticing that BN co-doping is found
to yield a weaker impact on the transmission probability when
compared with either B doping or N doping, \cite{doi:10.1021/nl901226s},\cite{PhysRevLett.102.096803} which can
be understood as a compensation effect of the local potential
distribution of both atoms in the dimer. \cite{PhysRevB.81.193411}

We now scrutinize the quantum transport properties of the
energetically most favorable configurations (adjacent B and
N atoms) and compare this type of GNR modification with
previous studies based on the total or partial suppression of the
$p_z$ orbital of some C atoms as a consequence of the chemical
attachment of the functional groups for the ribbon surface.
For oxygen atoms in an epoxide configuration,
\cite{ISI:000297143300092} the atomic orbitals of two consecutive carbon atoms adopt an intermediate
hybridization between $sp^2$ and $sp^3$. This results in a weak
conductance damage but substantial electron-hole asymmetry
of the resulting conductance profiles when considering wide
ribbons and low epoxide densities. For a complete $sp^3$ hybridization of two opposite C atoms in the hexagonal lattice, \cite{ISI:000268138600005} the hyperconjugation of the graphitic network is preserved,
although the presence of antiresonant states in the first plateau
dramatically decreases the transmission ability of the ribbon,
turning it into a strong insulator for the small density of
grafted atoms (such as H and OH groups) in agreement with 
\cite{ISI:000275901000019}. In sharp contrast, BN co-doping appears to impact
less markedly on the electronic conductance properties since
mean-free paths remain on the order of the micrometer, up to
the largest considered BN-co-doping density.

\begin{figure}[htp]
 \centering
 \includegraphics[width=0.45 \textwidth]{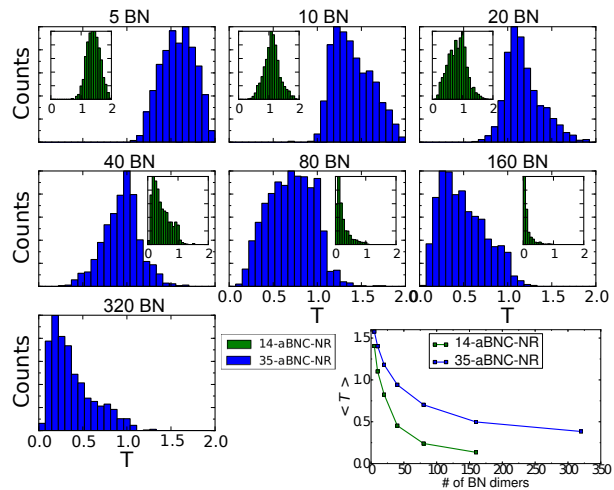}
 \caption{ Transmission histograms for BN codoped 35-aBNC-NR (main frames, blue) and 14-aBNC-NR (insets, green) with B and N atoms at first neighbouring positions randomly located in $L=2\mu$m long GNRs. The number of BN dimers in each transmission distribution is 5, 10, 20, 40, 80, 160, and 320. The evolution of the averaged transmission with respect to the number of BN dimers for each group of statistical distributions is also shown in the line plot graph.}                                                                                                                                                                                                                                                                                                                                             
 \label{FIGhisto}
\end{figure} 

Figure \ref{FIGhisto} shows the transmission distribution $T_{s(i)}(E)$ (with $s_{(i=1...N)}$ as a given disordered sample) for BN-co-doped 35-aBNC-NR (blue) and 14-aBNC-NR (green) at the selected energies with respect to the Fermi levels of 0.3 and 1 eV,
respectively. In all cases, B and N atoms are adjacent and are
randomly distributed along
$\sim 2\mu$m-long ribbons. For each
histogram, the statistical distribution is performed over 1000 random configurations. The line graphs $\langle T\rangle$ vs the number of dimers give the evolution of the averaged transmission decay for an increasing BN-co-doping rate in both types of ribbons with values ranging from 5 up to 320 BN dimers. For low BN-dimer density, the statistics is clearly seen to follow a Gaussian distribution around its sample average value $\langle T\rangle$, which remains larger than $G_0=e^{2}/h$, the quantum of conductance. The transport regime for 35-aBNC-NR (respectively, 14-aBNC-NR) with 10 BN-, 20 BN-, and 40 BN- (respectively, 5 BN-, 10 BN-, and 20 BN-) pair defects randomly distributed along the ribbon axis can, therefore, be
described by a quasiballistic or diffusive regime for which $\langle \frac{1}{T}\rangle\simeq 1/2+L/\xi$, if we denote $L$ the ribbon length and $\xi$ the localization length. \cite{PhysRevLett.87.116603} This allows us to estimate $\xi\sim$ 9 $\mu$m, $\xi\sim$ 4.4 $\mu$m and $\xi\sim$ 2.75 $\mu$m, respectively, for 5 BN-, 10 BN- and 20 BN- pair defects for the 14-aBNC-NR, whereas for the 35-aBNC-NR the values range from $\xi\sim $9$\mu$m down to $\xi\sim $3.5$\mu$m with varying BN dimer number from 10 up to 40. In all cases, the transmission and the localization length are found to roughly decay as $\sim 1/n_{BN}$ ($n_{BN}$ the number of pairs), in agreement with the Fermi golden rule.\cite{ISI:000252018300001}  

\begin{figure}[htp]
 \centering
 \includegraphics[width=0.45 \textwidth]{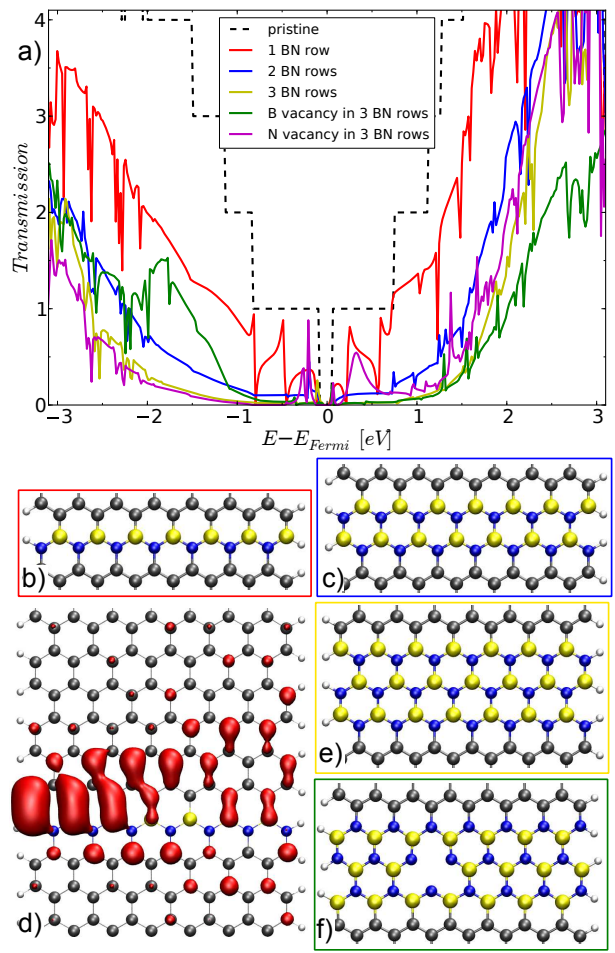}
 \caption{ a) Transmission coefficients of a GNR with BN domains covering the entire ribbon width. The various models considered for the calculation are given in the bottom panels, with varying number of aligned BN rows from 1 (b), up to 2 (c) and 3(e), and in presence of B monoatomic vacancy (f) for the 3-rows structure. The transmission along the ribbon decreases dramatically when two aGNRs are bridged by very few numbers of BN rows, although the presence of B (N) vacancies is shown to decrease (increase) the conductivity in the electron (hole) side. d) shows the charge distribution corresponding to the resonant state at -0.5 eV of the configuration in b). Yellow, blue, and gray spheres represent boron, nitrogen and carbon atoms, respectively. }
 \label{FIG2}
\end{figure}

When $\langle T\rangle \sim G_0$, a crossover is observed and the distribution widens to eventually becomes log-normal for $\langle T\rangle\ll G_0$. This behavior turns out to fully agree the universal conductance distributions derived for disordered systems \cite{PhysRevLett.87.116603}, as also found for disordered semiconducting doped nanowires \cite{PhysRevLett.99.076803,ISI:000252018300001}. The transmission fluctuations remain $\sim G_0$ whatever $\langle T\rangle$ as predicted by the Universal Conductance Fluctuations theorem. This happens to be the case for the 14-aBNC-NR ribbon with
80 BN and 160 BN pairs for which the conductance values and the obtained log-normal distribution (see Figure \ref{FIGhisto}) clearly point towards a localized regime defined by  $\langle lnT\rangle\sim -L/\xi$. We obtain $\xi$=20, 1.2, and 0.8 $\mu$m for 40 BN, 80 BN, and 160 BN pairs, respectively. The 35-aBNC-NR exhibits stronger robustness to disorder as evidenced by the distribution
of the transmission coefficients, which evolves more slowly to
a log-normal distribution, being only clearly visible for 320 BN
dimers and an average transmission of $\langle T\rangle=0.38$. 

\begin{figure*}[htp]
 \centering
 \includegraphics[width=0.95 \textwidth]{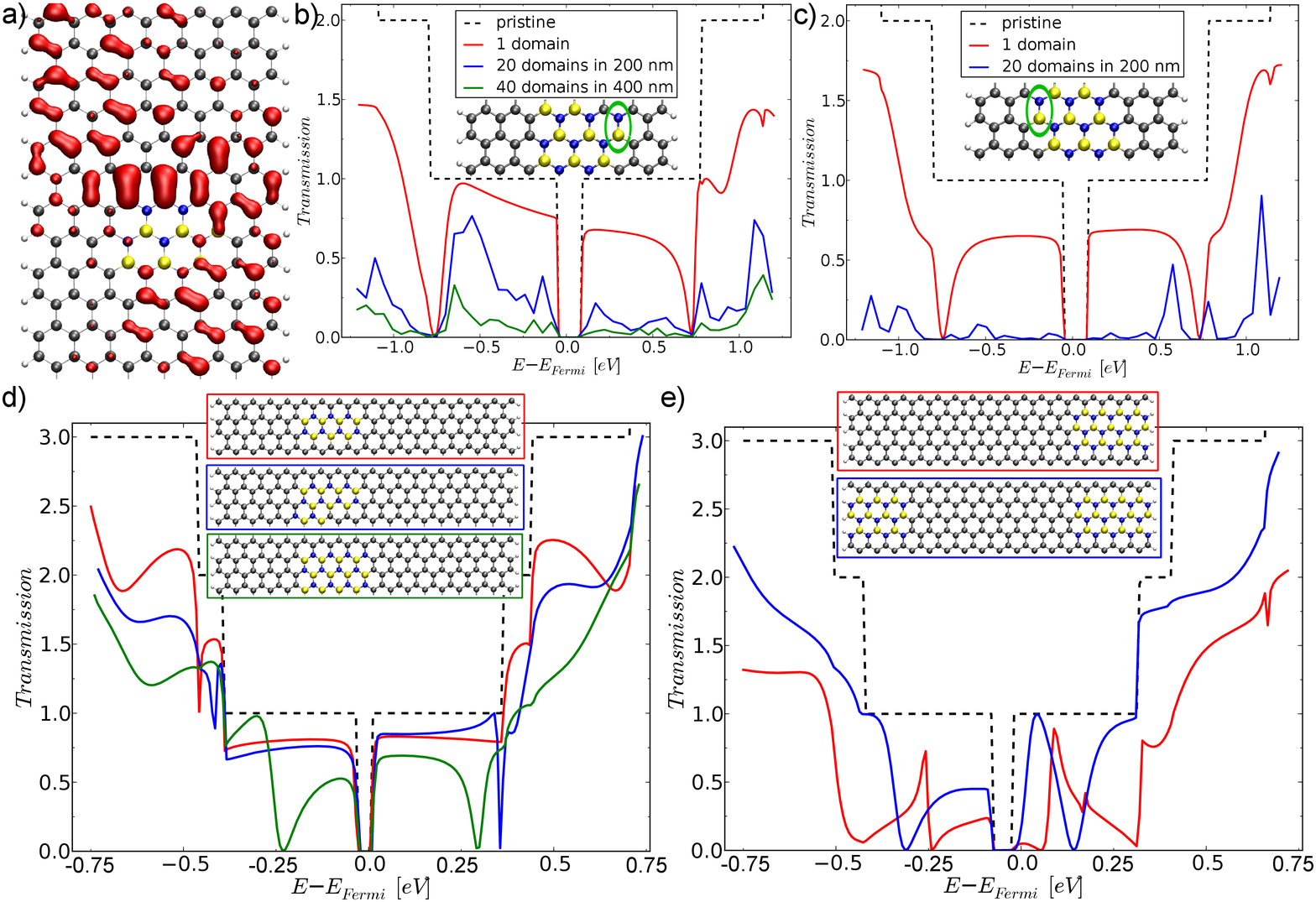}
 \caption{a) Electronic charge distribution for the antiresonant state at -0.85 eV corresponding to the configuration shown in the inset of b) and energy band diagram in Figure \ref{FIG0}-c). b) Averaged transmission for different numbers of domains randomly distributed along 2 micrometer of the 14-aBNC-NR with the BN domains represented in the inset. c) Same information for the a14-aBNC-NR with a different BN domain (as depicted in the inset). d) and e) show the calculated transmission coefficients for 35-aBNC-NR with BN domains or various size and shapes embedded in different regions of the ribbon. The frame color of each schematic representations inset corresponds with the color of the curves in the plots. Dashed lines give the transmission for the BN free ribbons. Yellow, blue, and gray spheres represent boron, nitrogen and carbon atoms, respectively.}
 \label{FIG3}
\end{figure*} 

We next investigate the effect on the electron transport of
BN barriers of different lengths and bridging two semi-infinite
graphene nanoribbons. The main results are summarized in Figure \ref{FIG2}. Firstly, a single row of BN pairs covering the whole ribbon width [Figure \ref{FIG2}-b)]  introduces significant backscattering throughout the considered energy spectrum with a noticeable electron-hole asymmetry. The difference between electron and hole conduction actually increases with Fermi energy in contrast to the prior results of single BN pairs distributed in a random fashion. Additionally, different from the previous case, many resonant states are observed in the first transmission plateau, which retain a high transmission probability. Increasing the number of rows up to 2 and 3 yields a dramatic
suppression of charge transport over a large energy window
of about 2 eV. This transport gap remains smaller than the
energy gap of bulk BN but clearly indicates that the BN barrier
acts as an insulating interface, which efficiently filters charge
transport. Tunneling effects only persist at higher energies as
seen in Figure \ref{FIG2}-a). Figure \ref{FIG2}-f) shows the efect of a monoatomic vacancy\cite{PhysRevLett.102.195505,Pochet2012} introduced in the interface formed by three consecutive BN
rows. The presence of both B and N vacancies is found to
modify the resonant tunneling regime at specific energies
by improving or reducing the transmission probability. See,
for instance, the enhanced transmission due to a B vacancy
in the energy range from -1 down to -2 eV. Similarly, a
N monovacancy in the three BN-row barriers is found to
enhance the electron transmission probability in the entire
conduction band with respect to the defect-free case. This
suggests that both B and N defects in BN patches might
create extended states in the domain that can contribute to the
ribbon conductivity. All these results show a large tunability
of transmission coefficients with the morphology of the BN
patches embedded within a graphene ribbon structure. One
notes that the presence of resonant states and antiresonances
[as illustrated in Figure \ref{FIG2}-d) and Figure \ref{FIG2}-a) for a chosen energy] for
which total reflection or transmission can develop, has been
well described for individual dopants in graphene nanoribbons in Ref. \cite{PhysRevLett.84.2917}.

We now focus on the description of electronic quantum
transport properties of GNRs with embedded BN domains
composed of an equal number of B and N atoms that remain
segregated in the carbon matrix. By tuning the compositional
range of the B, N, and C precursors, it is possible to
experimentally synthesize layers of graphene with different
amounts and sizes of BN domains,
 \cite{ISI:000276953500024} which always segregate
inside the graphene layers. Here, we restrict our calculations
to small width aBN-NRs to better assess the sensitiveness of
electronic transport to both disorder and BN-domain shapes
as seen in Figure \ref{FIG3}. The two cases studied only differ in the position of a single BN pair, represented with a green oval in the insets of Figure \ref{FIG3}-b) and Figure \ref{FIG3}-c). The transmission for
a single BN patch already shows substantial differences in
the first plateau and at higher energies. These differences are
further amplified by studying disordered hybrid ribbons of up
to 2-$\mu$-long. If 20 BN domains are enough to fully suppress
the conductance using a random distribution of BN domains as shown in Figure \ref{FIG2}-c), a stronger robustness is obtained for
the slightly different domain in Figure \ref{FIG3}-b) (inset). This situation
truly leads to insulating ribbons for system sizes on the order
of a few hundreds of nanometers.

The last structures investigated are large-width hybrid
ribbons (35-aBNC-NR) with various BN-domain sizes and
shapes as illustrated in the insets of 
Figure \ref{FIG3}-d) and -e). Here, we note that, by increasing the ribbon width and the
BN domain proportionally, we obtain transmission curves
similar to those of 14-aBNC-NR.
Figure \ref{FIG3}-d) shows how
weak structural changes can either let the transmission profile
unchanged or on the contrary, introduce new resonant states
and marked conductance suppression at the corresponding
resonant energies. A slightly different situation, shown in
Figure \ref{FIG3}-e), yields large variations in the conductance especially
in the first plateau for which the profile becomes even more
inhomogeneous. Indeed, we see a wide variability of transmis-
sion profiles, which is monitored by the precise BN-domain
shape and distribution across the ribbon structure. In realistic
BN-modified GNRs, a wide variability of BN patchworks in
terms of position, size, and shape are also expected, which
can be used to engineer the transport regimes unveiled in this
paper. The final transmission profile must be considered as the
result of a random combination of different BN domains so
that the quantum-mechanical details related to a specific BN
geometry is diluted in the multiple backscattering phenomena
and in the final average over a large ensemble of modified
ribbons.

\section{\label{SUMMARY}SUMMARY}
In conclusion, based on a combined approach of first-principles calculation and electronic-transport properties analysis, we have shown that the difficulties in opening a band gap
in graphene may be overcome by embedding BN domains
in the $\pi$-conjugated carbon network. We have analyzed
the electronic quantum transport characteristics of graphene
nanoribbons modified with embedded BN domains randomly
distributed along 2 $\mu$m by varying the size and position of the
BN domains. A wealth of transmission fingerprints, ranging
from energy-filtering interfaces or asymmetric electron-hole
transmission profiles to complete suppression of conductance,
have been found. The incorporation of foreign $sp^2$-hybridized
atoms with ionic characters into the graphitic network is not
transparent to the electronic transport. A crossover from the
ballistic to the localized regime was clearly evidenced by
the evolution of the statistical distribution of transmission
coefficients from a Gaussian to a log-normal shape for the
dilute BN-pair doping case, suggesting that a low-co-doping
rate could be enough for the transmission to enter the
localization regime. For larger BN-domain sizes, more insulating behaviors were obtained from asymmetric electron-hole 
conductance degradation to enlarged mobility gaps. Finally, 
complete insulating BN barriers to the current flow have 
been described, and the presence of resonant propagating 
states through defects (vacancies) inside the BN domain has 
been identified, complementing the rich variety of possible 
transmission phenomena and interference effects. 

This collection of results suggests a possible co-integration 
of a large quantity of complementary low-nanoscale devices 
based on the mixing of the local quantum properties of 
pristine BN and graphene. Indeed, whereas, clean or weakly 
BN-co-doped graphene maintain very good conduction capability and could serve as interconnects, BN patches with 
varying sizes embedded within the graphene matrix offer larger
tunability of current flow, owing to the induced mobility gaps
or insulating barriers.

% \acknowledgment

\section{\label{Acknowledgments}Acknowledgments}
This research used the resources of the National Center for
Computational Sciences at Oak Ridge National Laboratory,
which is supported by the Office of Science of the US Depart-
ment of Energy under Contract No. DE-AC05-00OR22725.
We are also grateful for support from the Center for Nanophase
Materials Sciences (CNMS), sponsored at Oak Ridge National
Laboratory by the Division of Scientific User Facilities, US
Department of Energy.

% \bibliography{biblio}{1}  
% \bibitem{AjayanBNtrans}
% Song,L. ;  Balicas,L.;  Mowbray,D. J.; Capaz, R. B. ; Storr,K.; Ci, L.;  Jariwala,D. ; Kurth, S.; Louie, S.G.;Rubio,A. ;  Ajayan,P.M., Phys. Rev. B. (submitted), arXiv:1105.1876
  
\end{document}